# Modulation of magnetic property of double-perovskite La$_2$MnCoO$_6$ films by controlling *B*-antisite disorder

R. C. Sahoo[1], S. Zhang[1,], Y. Takeuchi[1], M. Kitamura[2], K. Horiba[2,3], H. Kumigashira[2,3,4], O. Sakata[3,5], T. Soma[1], and A. Ohtomo[1,3,*]

[1]Department of Chemical Science and Engineering, Tokyo Institute of Technology, Tokyo 152–8552, Japan

[2]Photon Factory, Institute of Materials Structure Science, High Energy Accelerator Research Organization (KEK), Tsukuba 305–0801, Japan

[3]Materials Research Center for Element Strategy (MCES), Tokyo Institute of Technology, Yokohama 226–8503, Japan

[4]Institute of Multidisciplinary Research for Advanced Materials, Tohoku University, Sendai 980–8577, Japan

[5]Synchrotron X-ray Group, Research Center for Advanced Measurement and Characterization, National Institute for Materials (NIMS), Sayo, Hyogo 679–5148, Japan

* ohtomo.a.aa@m.titech.ac.jp

**[Abstract]**

The predicted physical properties of double perovskites have usually been compromised by their intrinsic *B*-antisite (*AS*) disorder effect, but the relationship between them is still unclear for the $La_2MnCoO_6$ (LMCO) system. This study focuses on controlling the *AS* disorder and quantitatively reveals correlations with magnetic and electronic states in epitaxial LMCO films grown on (111) $SrTiO_3$ substrates. The *AS* fraction was precisely controlled by tuning the growth conditions as evaluated from saturation magnetization and x-ray reflection profiles. The saturation magnetization at 5 K decreased linearly from 6 $\mu_B$ per formula unit by 50% as the *AS* fraction increased from 0 (perfectly ordered) to 0.5 (fully disordered). The x-ray absorption spectroscopy revealed that valence states of Mn and Co were 4+ ($3d^3$) and 2+ ($3d^7$), respectively, regardless of *AS* fraction. A local-spin-moment model is proposed to explain the net magnetizations with and without *AS* disorder and Mn/Co valence states, suggesting that only the rearrangement of local spin moment of Co at the antisite plays a key role in tuning the magnetism in this LMCO system.

**[Introduction]**

The periodicity and symmetry of crystal structures have a big impact on the physical properties of a strongly correlated electron system. A subtle change of imperfection and/or symmetry breaking of the crystal structure have been utilized to tune physical properties such as ferromagnetism, charge transport, competing magnetic interactions, magnetic exchange interactions, and thermal conductivity [1–5], depending on targeted applications. The double-perovskite-type $A_2BB'O_6$ (where $A$ and $B/B'$ are rare-earth (and/or lanthanide) and transition-metal elements, respectively) is a class of the strongly correlated systems whose crystal structures, especially the periodicity of the $B$-site cations, strongly influence their magnetic and electronic properties [6]. In such systems, edge-shared $BO_6$ and $B'O_6$ octahedra often resemble the NaCl-type structure. But most systems inherently exhibit antisites of either $BO_6$ or $B'O_6$ octahedra, commonly known as $B$-antisite ($AS$) disorder, which is one of primary factors to regulate the magnetic properties. In addition, a moderate $AS$ disorder often leads to complexity in exchange interactions among $B$-O-$B'$, $B$-O-$B$, and $B'$-O-$B'$ pairs, each of which generally obeys the Kanamori and Goodenough rule [7,8]. Therefore, the ground states can be manipulated by tuning *AS* disorder in the double perovskites. However, there is no systematic and universal relationship between the *AS* disorder and the physical property, particularly the magnetic property, in such perovskite systems.

Among the well-studied double perovskites, La$_2$Mn$B'$O$_6$ ($B'$ = Fe, Co, and Ni) is of great interest for investigating underlying magnetic couplings in oxides [9–11]. A number of studies on La$_2$MnCoO$_6$ (LMCO) in the form of polycrystals and thin films revealed their ferromagnetic-insulating nature near at room temperature [12–15]. The ferromagnetic (FM) behaviors with a large magnetization of ~6 $\mu_B$ per formula unit (f.u.) was observed below 225 K due to the superexchange interaction between Co$^{2+}$ and Mn$^{4+}$ ions [16]. Partially *AS*-disordered LMCO exhibited multiferroicity and relaxer-like dielectric behavior, making it a rare example of an intrinsic multifunctional system [17,18]. On the other hand, a thermodynamically stable phase of LMCO is the ordered double perovskite, which arises from distinct ionic radii and different valence states for Co$^{2+}$ and Mn$^{4+}$ ions. However, it has not been explored yet to grow near-perfectly disordered LMCO single crystal, which prevents us from revealing *AS*-disorder effects on magnetism. Therefore, correlation between degree of *AS* disorder and magnetism in LMCO has not been understood yet.

In recent, the precise control of complex-oxide epitaxy can be realized thanks to the recent advances in deposition techniques [19]. In particular, naturally disordered double-perovskite La$_2$MnFeO$_6$ (LMFO) has been transformed to ordered phases by high-temperature growth with pulsed-laser deposition (PLD) [20]. It is noticeable that both identical ionic radii (64.5 nm) and valence states for Mn$^{3+}$ and Fe$^{3+}$ with high-spin configurations do not preclude spontaneous *B*-site ordering. In this case, the correlation between degree of *AS* disorder and magnetism is not pronounced owing to the presence of LaMnO$_{3+\delta}$ domains, in which local spin moment of the Mn 3*d* state ferromagnetically align. Furthermore, local spin moments of the Mn 3*d* and Fe 3*d* states align antiparallel to each other. Because of ferromagnetic LaMnO$_{3+\delta}$ domains, this ferrimagnetic ground states survive even in samples with high *AS* disorder. It is a surely new challenge to transform spontaneously ordered double-perovskite LMCO to disordered phases and to reveal how ferromagnetic LaMnO$_{3+\delta}$ domains influence net magnetization.

In this paper, we report on the synthesis of *AS*-disorder controlled double-perovskite LMCO films and their magnetic properties as a function of *AS* fraction from 0 (perfectly ordered) to 0.5 (fully disordered). The *AS* fraction was examined by x-ray diffraction (XRD) with synchrotron radiation. The magnetization behaviors were investigated with a vibrating sample magnetometer (VSM). We found that the estimated saturation magnetization ($M_s$) and the *AS* fraction shared a linear relationship with negative correlation. The x-ray absorption spectroscopy (XAS) was used to confirm the valence states of Co and Mn cations, showing no dependence of the valence states on *AS* disorder. Finally, a schematic spin texture was used to explain the link between $M_s$ and *AS* fraction. The successful growth and systematic behavior of magnetic properties of the *AS*-disorder controlled LMCO provide us useful information in a framework of an important class of magnetic double perovskites, La$_2BB'$O$_6$ ($BB'$ = Mn, Fe, Co, and Ni).

**[Experiments]**

The LMCO films were grown by using PLD equipped with an ultra-high-vacuum chamber, a KrF excimer laser (248 nm, 10 Hz, 1.2 J/cm$^2$), and *in-situ* reflecting high-energy electron diffraction (RHEED). The dynamic growth process was monitored by RHEED. For the ablation process, a stoichiometric polycrystalline LMCO target was synthesized by a conventional solid-state reaction method. The films with thickness ranging from 45 to 110 nm were grown on TiO$_2$-terminated (111) SrTiO$_3$ (STO) substrates (step-and-terrace like surface) in oxygen pressure ($P_{O2}$) ranging from $1.0 \times 10^{-7}$ to 1.0 Torr using a continuous flow of pure O$_2$ (6N) at growth temperature ($T_g$) ranging from 400 to 800 °C. After the growth, $T_g$ was decreased to room temperature at a rate of ~50 °C/min with keeping $P_{O2}$ constant.

The epitaxial structure of as-grown films was verified from profiles of 2*θ*-*ω* scans and *ω*-scan rocking curves and reciprocal space mapping (RSM) using a laboratory XRD apparatus with Cu *K*α radiation (*λ* = 1.5405 Å). In addition, synchrotron-radiation XRD measurements were carried out at the undulator beamline of BL15XU in SPring-8 to disclose degree of *AS* disorder. The photon energy of the incident x-ray was set to 15 keV. The intensity of obtained *hhh* reflections was corrected by integrating rocking-curve profiles measured along the [11$\bar{2}$] direction of the STO substrates. Both temperature and field-dependent magnetization properties were measured by using a VSM equipped with a physical property measurement system (PPMS-VSM, Quantum Design). The room-temperature XAS measurements were carried out at the undulator beamline of BL2A in Photon Factory, KEK.

**[Results and Discussion]**

A clear relationship between the growth conditions ($T_g$-$P_{O2}$) and $M_s$ was found for LMCO films. Figure 1(a) depicts a $T_g$-$P_{O2}$ diagram mapping the results of structural characterization and $M_s$ measured at 5 K. The high-order films were obtained reproducibly around $T_g$ = 700 °C and $P_{O2}$ = $10^{-5}$ Torr (marked as Sample A), and *AS* fraction reached 0.018(5) as will be discussed later. Note that volume fraction of the ordered phase as large as 96% (= 1–2*AS*), reflecting spontaneous ordering tendency. From previous studies, it is known that high-temperature growth (*c.a.* 1000 °C) is required to achieve ordered phases of naturally disordered double perovskites films (such as LMFO) [20–23]. The intermediate-$T_g$ condition for LMCO is in striking contrast to them. Moreover, the crystallinity of the as-grown films gradually deteriorated in going away from this sweet spot to high-$P_{O2}$ or high-$T_g$ region. In particular, an amorphous phase appeared at 500 °C and $10^{-1}$ Torr (Sample D), while secondary phases were observed in a wide range including one at 800 °C and $10^{-3}$ Torr (Sample C). On the other hand, disordered phases with no secondary phase were obtained in a low-$T_g$/low-$P_{O2}$ region. No significant degradation of crystallinity was observed even at 400 °C (Sample B). In fact, a clear and long-lasting RHEED intensity oscillations were observed for not only Sample A but also Sample B, indicating a layer-by-layer growth [Fig. 1(b)]. In contrast, a rapid intensity damping was observed during the growth of Sample D, reflecting amorphous phase and/or a three-dimensional growth mode (not shown).

The out-of-plane $2\theta$-$\omega$ profiles for Samples A–D are shown in Fig. 1(c). The NaCl-type ordering of Mn/Co pairs was evidenced from superlattice (SL) reflections of odd *hhh* observed at $2\theta \sim 20°$ and $60°$ for Sample A. Because of similar atomic scattering factors of Mn and Co, the intensity of SL reflections is relatively weak. Interestingly, these SL reflections are also observed along with 111 and 222 reflections of CoO as a secondary phase in the $2\theta$-$\omega$ profile for Sample C. However, SL reflections for the other films are not found in the $2\theta$-$\omega$ profiles. As for Sample D, even reflections from a fundamental perovskite lattice were absent, suggesting amorphous phase. We selected high-order (Sample A) and low-order (Sample B) films for measuring $\omega$-scan rocking curve profiles of the LMCO 222 reflection. As shown in Fig. 1(d), sharp peaks with small full width at half maximum (0.12 and 0.086° for Samples A and B, respectively) confirmed excellent crystallinity of both films. The epitaxial relationship has been verified through RSM measurement around the asymmetric 330 reflection of the STO substrate. The position of asymmetry reflections confirmed that all the films studied here were coherently grown on STO (111) [see Fig. 1(e) for the example of Sample A]. We estimated the average lattice constant, $a_{av} = V_{av}^{1/3}$, where $V_{av}$ is the unit-cell volume being equal to $6\sqrt{3}d_{222}(d_{11\bar{2}})^2$. The estimated $a_{av}$ of Sample A was ~3.909 Å, which was longer than those of polycrystalline LMCO used for the PLD target ($a_{av}$ ~ 3.886 Å) and reported single-crystalline LMCO ($a_{av}$ ~ 3.862 Å) [24]. Meanwhile, $a_{av}$ of Sample B was ~3.940 Å, much longer than that of ordered Sample A. Note that these $V_{av}$ have been used to estimate accurate $M_s$ of the films investigated in this study.

Synchrotron-radiation XRD measurements were performed to further investigate fraction of *AS* disorder in our LMCO films. For example, Figure 1(f) represents a symmetric reflection profile for high-order Sample A. We used the intensity of the fundamental LMCO 444 reflection to normalize the observed intensity of the other reflections with *h* from 1 to 7. The SL reflections were intense enough to perform fit to a crystal model, where *AS* and out-of-plane displacement of $La^{3+}$ and $O^{2-}$ were considered to be variables [22]. We have also succeeded in observing SL reflections for lower-order films by using high-flux synchrotron radiation and estimating *AS* fraction. A satisfactory agreement between calculated and measured intensity was found for all films investigated in this study. As described earlier, *AS* fraction of Sample A was 0.018(5) and corresponding volume fraction of the ordered phase was as large as 96%. As for Sample B, *AS* fraction substantially increased to 0.32(6) and volume fraction of the ordered phase was only 35%. In the following sections on magnetism, we also refer to *AS* fraction [0.061(5) and 0.49(9)] of additional samples grown under the condition identical to those for Samples A and B, respectively.

To reveal magnetic properties and their dependence on *AS* fraction, temperature- and field-dependent magnetizations were measured for a number of films with various *AS* fraction. Figure 2(a) shows the in-plane ($H$ // $[1\bar{1}0]$) field-dependent magnetization (*M-H*) measured at 5 K for Samples A and B. Well-defined hysteresis loops were observed after subtracting the diamagnetic signal of the STO substrate for all the samples considered in this study. We found that $M_s$, remanent magnetization, and coercive field monotonically decreased with increasing *AS* fraction. The $M_s$ was estimated to be 5.8 $\mu_B$/f.u. for Sample A with *AS* ~ 2% and ~4.0 $\mu_B$/f.u. for Sample B with *AS* ~ 33%.

The estimated $M_s$ for Sample A agrees well with the theoretically calculated $M_s$ of LMCO with the existence of $Mn^{4+}$ [$3d^3$ ($t_{2g}^3$ ↑); $S = 3/2$] and $Co^{2+}$ with a high-spin configuration [$3d^7$ ($t_{2g}^3$ ↑ $t_{2g}^2$ ↓ $e_g^2$ ↑); $S = 3/2$] ions [25]. The FM ordering of these local spin moments ($3d^3$ ↑ $3d^7$ ↑; $S_{total} = 3/2 + 3/2$) leads to spin only moment of 6 $\mu_B$/f.u. In contrast, other samples exhibit lower $M_s$ due to *AS* disorder. The lower or even higher $M_s$ than 6 $\mu_B$/f.u. are previously discussed in terms of not only *AS* disorder but also mixed valence states of $Mn^{4+}/Co^{2+}$ and $Mn^{3+}/Co^{3+}$ [24,26,27]. In this study, we found a simple scenario for lower $M_s$ in our disordered films, as will be discussed later. Moreover, hysteresis loops of Samples A and B indicate the contribution of two phases having different coercive fields. This behavior was absent and only the component with high coercive field was observed for fully disordered sample with *AS* ~ 0.5. Such a step-like behavior was previously observed for bulks and thin films, and is attributed to the coexistence of a spin-glass-like state with FM state or two FM states [24,27]. We also observed spin-glass-like behaviors as described below.

Figure 2(b) shows the temperature dependence of magnetization ($M$-$T$) under field-cooled (FC) and zero-field-cooled (ZFC) modes for Samples A and B. A gradual increase in magnetization was observed for Sample A below 135 K, which was verified Curie temperature ($T_C$) from Curie-Weiss fitting in the high-temperature regime. The Sample B showed FM ordering below 100 K. The other samples showed similar behaviors with $T_C$ ranging from ~100 K to ~150 K. In contrast to $M_s$ described above, there was no significant dependence of $T_C$ on $AS$ fraction. The reported $T_C$ distributes in a wide range temperature from 230 K to 150 K [16,24,26,27]. In particular, a recent study on single crystal LMCO finds $T_C$ ~ 150 K [24]. From these results, our sample indicated somewhat lower temperatures for FM ordering. The relatively longer $a_{av}$ likely causes the lower $T_C$, although there was no systematic dependence of $T_C$ on $a_{av}$ among all the samples studied here. We observed a strong bifurcation between the ZFC and FC magnetization curves for all the samples below 100 K. These behaviors generally suggest the presence of multiple magnetic phases including spin-glass-like phase, and have been commonly observed for LMCO polycrystals and single crystals regardless of the degree of *AS* disorder and/or mixed valence states of Mn/Co [16,24,26].

In order to investigate roles of *AS* disorder and mixed valence states on magnetic properties of our samples, we performed XAS measurements at 300 K for Samples A and B. Figure 3(a) shows Mn 2*p* XAS spectra of both samples along with the reference spectra of $Mn^{4+}$ ($MnO_2$), $Mn^{3+}$ ($LaMnO_3$), $Mn^{2+}$ (MnO), and $Mn^{2+,3+}$ ($La_{0.7}Ce_{0.3}MnO_3$) [28]. These spectra indicate a number of similarities, including two strong peaks centered at ~645 eV and ~656 eV, respectively, corresponding to the $L_3$ and $L_2$ edges. Moreover, one can find that the spectral features of the $Mn^{4+}$ species in $MnO_2$ are very similar to these spectra, suggesting that $Mn^{4+}$ [$3d^3$ ($t_{2g}^3$)] is predominant in our samples regardless of the *AS* disorder. Figure 3(b) shows Co 2*p* XAS spectra along with the reference spectra of $Co^{4+}$ (calculated), $Co^{3+}$ [$LaCoO_3$; high- (HS), intermediate- (IS), and low-spin (LS) states], and $Co^{2+}$ (CoO) [29,30]. These spectra again have a number of similarities, including prominent peaks of the $L_3$ edge at ~778 eV. These spectral features are nearly identical to those of $Co^{2+}$ species, suggesting that $Co^{2+}$ with a high-spin configuration [$3d^7$ ($t_{2g}^5 e_g^2$)] is predominant, again regardless of *AS* disorder. Therefore, only $Mn^{4+}$ and $Co^{2+}$ species are responsible in our studied samples for interactions of local spin moments and resulting magnetism.

Based on the results described above, we plotted $M_s$ under 5 T at 5 K taken from *M-H* curves as a function of *AS* fraction [Fig. 4(a)]. The $M_s$ decreased linearly with increasing *AS* fraction, and least-square fitting yielded a relationship of $M_s$ = 5.88–5.77*AS*. Note that $M_s$ at *AS* = 0.061(5) and 0.49(9) are 5.5 and 3.0, respectively. These values were taken from additional samples grown under the condition identical to those for Samples A and B, respectively.

The observed linear relationship allows us to discuss possible arrangements of local spin moments at the *AS* defects. Figure 4(b) shows an atomic model of local spin moment at the cross section of pseudocubic units projected along the {*h*00} direction. Based on XRD, *M-H*, and XAS measurements, $Mn^{4+}$ and $Co^{2+}$ ions are arranged in a NaCl-type lattice, and local spin moments of $S$ = 3/2 at each site are parallel to the magnetic field. In the top panel, each $Mn^{4+}$ ion is surrounded by six $Co^{2+}$ ions and vice versa, representing the perfect Mn/Co ordering. When the position of a single Mn/Co pair is altered, partially disordered lattice with *AS* = 1/9 is created as shown in the bottom panel. In this case, a $Mn^{4+}_{Co}$ ($Co^{2+}_{Mn}$) ion at the site labelled A (B) is surrounded by six $Mn^{4+}_{Mn}$ ($Co^{2+}_{Co}$) ions. Although the antiferromagnetic superexchange of $Co^{2+}$-O-$Co^{2+}$ and $Mn^{4+}$-O-$Mn^{4+}$ dominate at both sites, contributions of ferromagnetic $LaMnO_{3+\delta}$ domains must to be considered at the site A. As a result, local spin moment of antisite $Mn^{4+}_{Co}$ ions remains intact and does not contribute to change in net magnetization. Therefore, only local spin moment of antisite $Co^{2+}_{Mn}$ ions reduces the magnetic moment from the rest of the ordered part by 6 $\mu_B$/f.u. Based on this model, change in $M_s$ with $AS$ fraction is predicted to follow $M_s = 6(1-AS)$ $\mu_B$/f.u., which is in excellent agreement with a linear relationship found experimentally.

Focusing on the atomic arrangement, local structures at the antisite *clusters* can be regarded as $LaMn^{4+}O_{3+\delta}$ and $LaCo^{2+}O_{3-\delta}$. It has been previously confirmed that both $LaMnO_3$ and $LaCoO_3$ have an antiferromagnetic ground state according to the Kanamori–Goodenough rule [31–33]. If disordered LMCO obeys this tendency, the antisite $Mn^{4+}_{Co}$ ions also contribute to the reduction of net magnetization as $M_s = 6(1-2AS)$ $\mu_B$/f.u. This model largely fails to explain the observed relationship between $M_s$ and $AS$ fraction. Another possibility is that only local spin moment of antisite $Co^{2+}_{Mn}$ ions is maintained. In this case, the experimental results could also be explained. However, it is well-known that LaMnO$_3$ is much more likely to exhibit FM ordering than LaCoO$_3$. In fact, the previous studies on LaMnO$_3$ demonstrated FM ordering when a small amount of excess oxygen and La deficiency were introduced [20,34]. As for disordered LMCO, ferromagnetism prevails at the antisite Mn$_{Co}$, whereas antiferromagnetism dominates at the antisite Co$_{Mn}$. Thus, each *AS* defect converts a Mn$^{4+}$/Co$^{2+}$ pair with a total moment of 6 $\mu_B$ (coupled ferromagnetically) with the rest of the sample into a pair carrying 3 $\mu_B$ (coupled antiferromagnetically). This simple model also works well for other disordered LMCO films examined in this study.

**[Conclusion]**

In summary, we have investigated the effects of *AS* disorder on magnetic properties of La$_2$MnCoO$_6$ epitaxial thin films grown on STO (111) substrates. The *AS* fraction was found to be highly sensitive to the growth conditions selected for the thin-film growth. Meanwhile, we established recipes for the layer-by-layer growth of both ordered and disordered LMCO films and presented them in a phase stability diagram. The high-order film exhibited the degree of Co/Mn order as high as 96% and $M_s$ ~ 5.8 $\mu_B$/f.u., while lower values were obtained for films with various *AS* fractions. The $M_s$ and *AS* fraction followed a linear relationship with negative correlation. For example, a fully disordered film had $M_s$ ~ 3 $\mu_B$/f.u., which was explained by a local-spin-moment model. The XAS spectra suggested tetravalent Mn and divalent Co ions with a high-spin configuration, irrespective of *AS* fraction. Through systematic characterization of the studied samples having a wide range of *AS* fraction, we have clearly shown that the magnetic ground state of LMCO double perovskite depends only on the *AS*-disorder-driven spin texture of $Co^{2+}$ ions. In general, our study will contribute to understanding the functional properties of various double perovskites driven not only by types of *B*-site cations, but also by deliberately introduced *AS* disorder.

**[Acknowledgments, Funding]**

This work was partly supported by MEXT Element Strategy Initiative to Form Core Research Center (Grant No. JPMXP0112101001) and a Grant-in-Aid for Scientific Research (No. 20K15169 and No. 21H02026) from the Japan Society for the Promotion of Science Foundation. The synchrotron XRD measurements were performed under the approval of the NIMS Synchrotron x-ray Station at SPring-8 (Proposals No. 2019B4701 and No. 2020A4750). The work at KEK-PF was done under the approval of the Program Advisory Committee (Proposals No. 2019G583 and No. 2018S2-004) at the Institute of Materials Structure Science, KEK. This research also was partly conducted using research equipment shared in MEXT Project for promoting public utilization of advanced research infrastructure (Program for supporting introduction of the new sharing system) Grant Number JPMXS0420900521. RCS is supported by Japan Society for the Promotion of Science for Scientific Research.

**[References]**

**[Figure Caption]**

**FIG. 1.** Growth and structural characterizations of LMCO films. (a) PLD growth conditions of LMCO films mapped in a $T_g$-$P_{O2}$ diagram. The markers indicate apparent sample information determined from XRD $2\theta$-$\omega$ scan, as the samples observed with and without SL reflections (○ and □, respectively), and/or with a secondary phase of CoO (×) or with an amorphous phase (△). Inset color mapping shows saturation magnetization of single-phase LMCO films measured at 5 K. (b) RHEED intensity oscillations for the specular beam during the growth of Samples A and B. (c) Out-of-plane XRD profiles for Samples A–D. The colors in (c) correspond to those in (a). (d) x-ray rocking curves of LMCO 222 reflections for Samples A and B. (e) Reciprocal space map taken around STO 330 reflections for Sample A. (f) Symmetric reflection profiles taken with synchrotron-radiation source (open circles). The bars indicate calculated intensities for each reflection. Both profiles are normalized by the peak intensity of the 444 reflection.

**FIG. 2.** Magnetic properties of LMCO films. (a) Magnetic hysteresis loops taken at 5 K

and (b) temperature dependence of FC (■) and ZFC (□) magnetization, measured under a magnetic field of 50 mT, for Samples A (red) and B (blue). For both (a) and (b), the magnetic field was applied parallel to the film surface.

**FIG. 3.** Normalized XAS spectra. (a) Mn 2$p$ XAS spectra taken at 300 K for Samples A (red) and B (blue). The spectra of $Mn^{2+}$ (MnO), $Mn^{3+}$ (LaMnO$_3$), $Mn^{2+,3+}$ (La$_{0.5}$Ce$_{0.5}$MnO$_3$), and $Mn^{4+}$ (MnO$_2$) are also shown as references [28]. (b) Co 2$p$ XAS spectra taken at 300 K for the same samples. The spectra of $Co^{2+}$ (CoO), $Co^{3+}$ HS, $Co^{3+}$ IS, $Co^{3+}$ LS (LaCoO$_3$), and $Co^{4+}$ (calculation) are also shown as references [29,30].

**FIG. 4.** Correlation between magnetism and *AS* disorder. (a) Saturation magnetization ($M_s$) measured under 5 T at 5 K as a function of *AS* fraction simulated from synchrotron XRD. (b) Model of local spin moment at a pair of antisite Mn$_{Co}$ and Co$_{Mn}$ ions, showing a net spin-moment compensation of 3$\mu_B$ per pair. Note that the arrangement of local spin moment at the Mn$_{Co}$ and Co$_{Mn}$ site is opposite to each other.

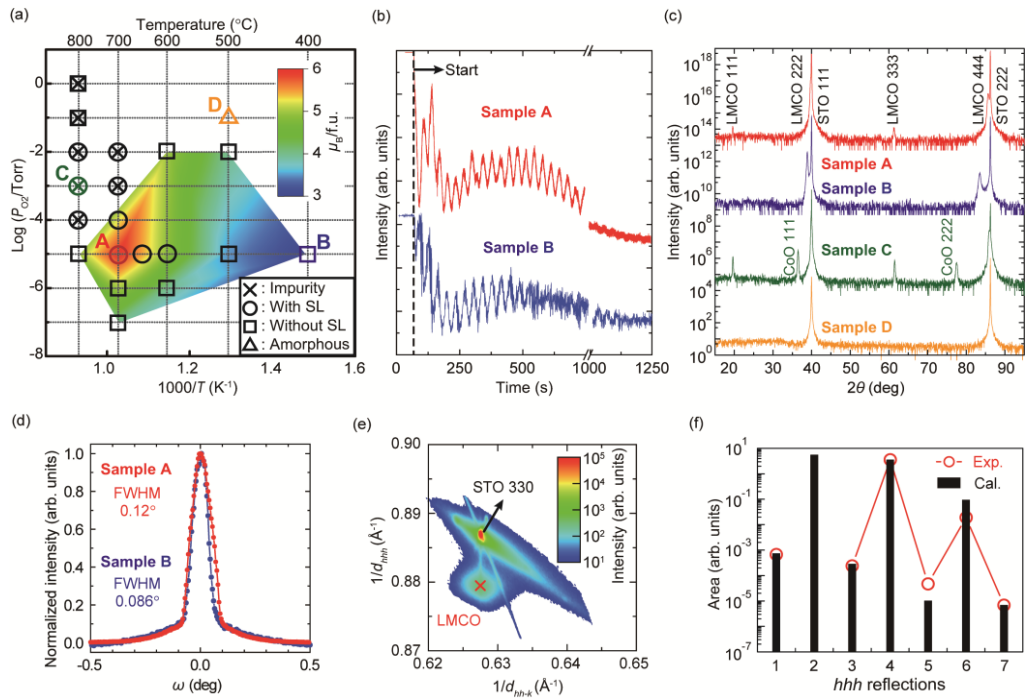

Figure 1 Sahoo *et al*.

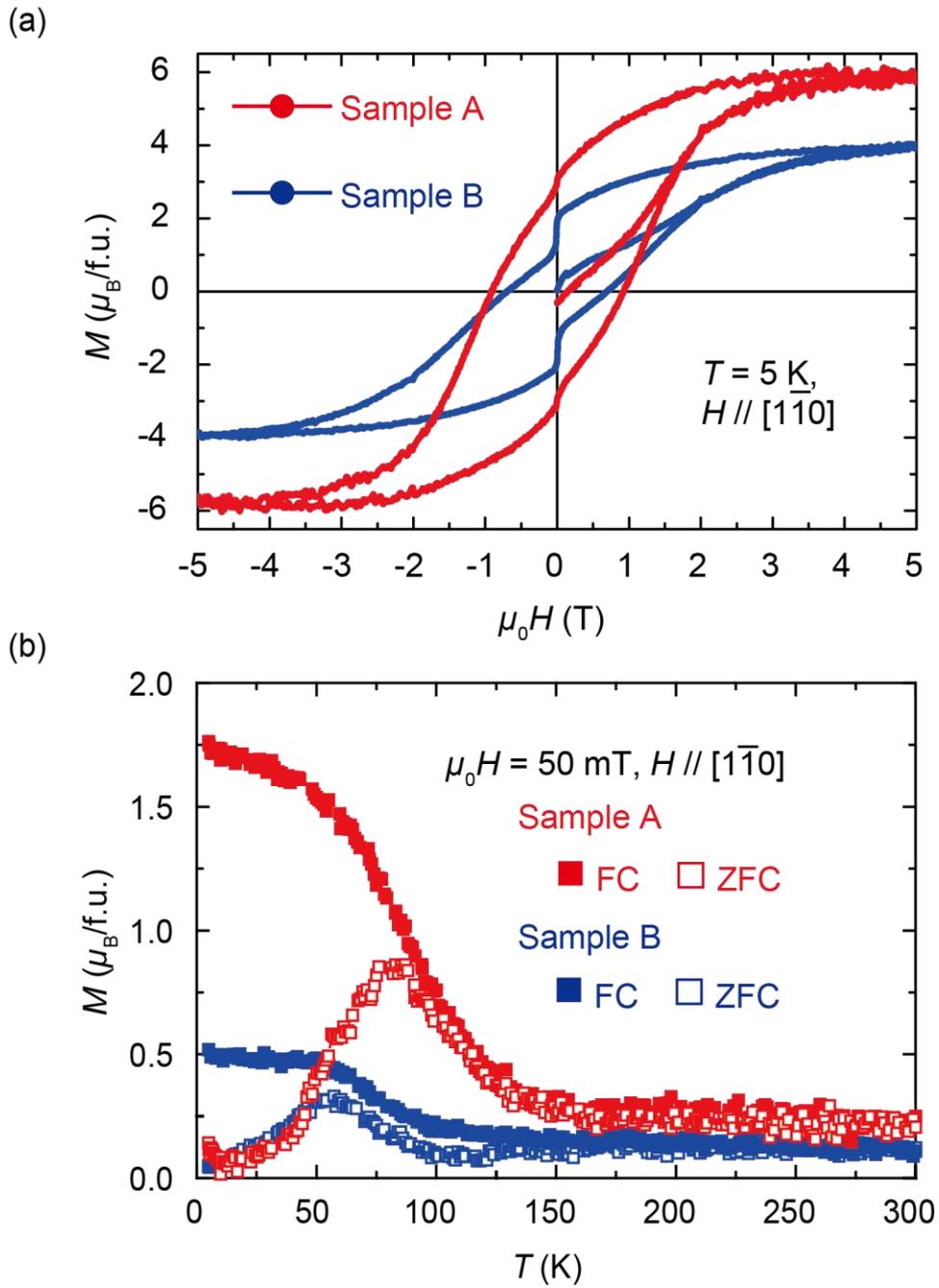

Figure 2 Sahoo *et al*.

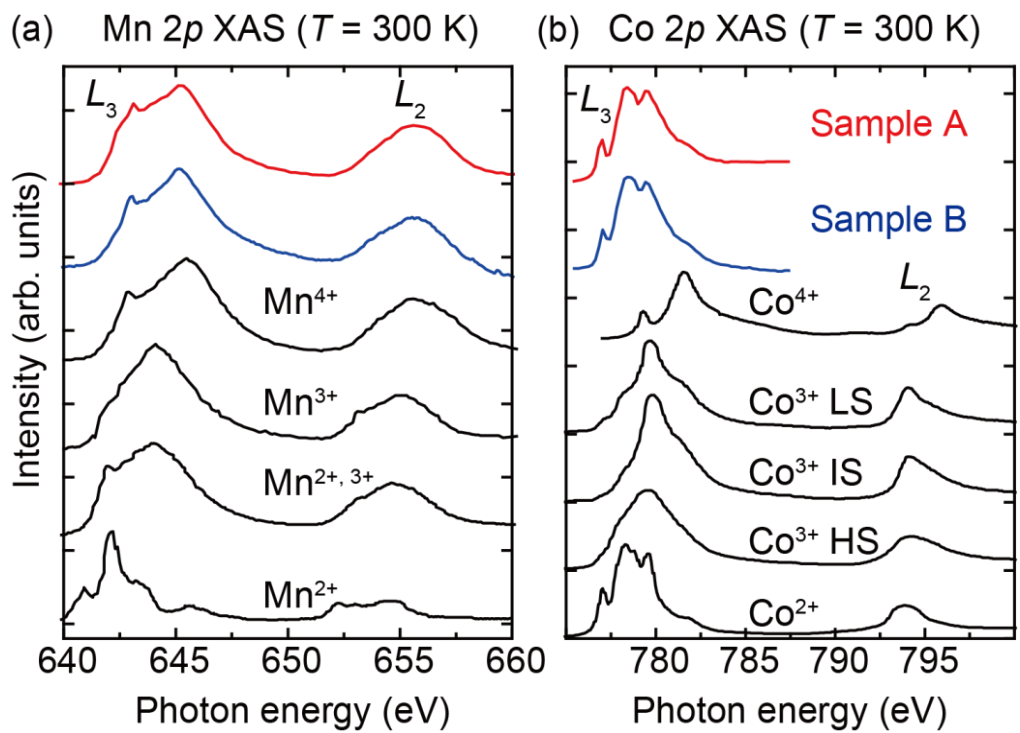

Figure 3 Sahoo *et al*.

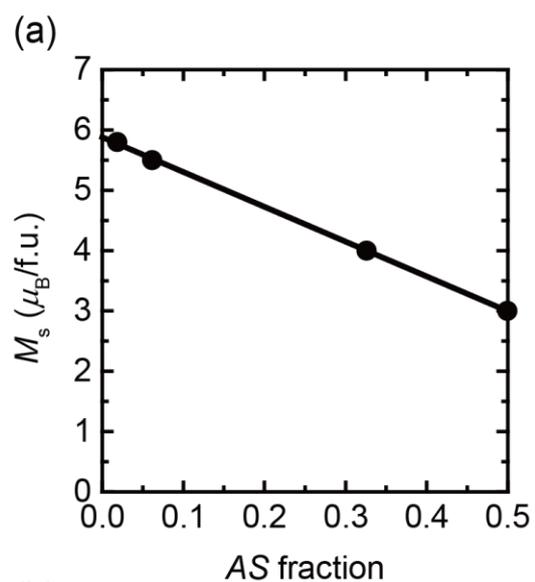
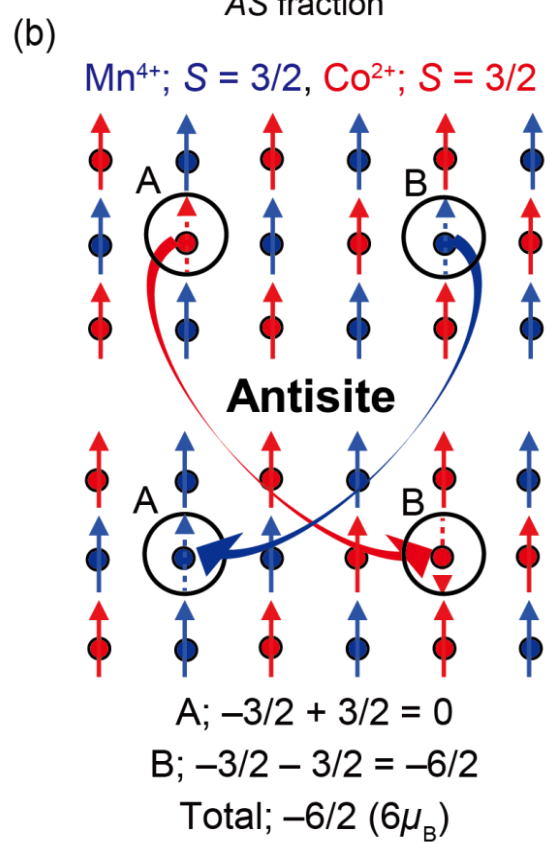

A; −3/2 + 3/2 = 0
B; −3/2 − 3/2 = −6/2
Total; −6/2 (6$\mu_B$)

Figure 4 Sahoo *et al*.